<Paper>

# Void-Defect Induced Magnetism and Structure Change of Carbon Material-III: Hydrocarbon Molecules

Norio Ota[1], Aigen Li[2], and Laszlo Nemes[3]

[1]Graduate school of Pure and Applied Sciences, University of Tsukuba, *1-1-1 Tennodai, Tsukuba-City Ibaraki 305-8571, Japan*
[2]Department of Physics and Astronomy, University of Missouri, *Columbia, MO 65211, USA*
[3]Research Center for Natural Sciences, Ötvös Lóránd Research Network, *Budapest 1519, Hungary*

Void-defect induced magnetism of graphene molecule was recently reported in our previous paper of this series study. This paper investigated the case of hydrogenated graphene molecule, in chemical term, polycyclic aromatic hydrocarbon (PAH). Molecular infrared spectrum obtained by density functional theory was compared with astronomical observation. Void-defect on PAH caused serious structure change. Typical example of $C_{23}H_{12}$ had two carbon pentagon rings among hexagon networks. Stable spin state was non-magnetic singlet state. This is contrary to pure carbon case of $C_{23}$, which show magnetic triplet state. It was discussed that Hydrogen played an important role to diminish magnetism by creating an SP3-bond among SP2-networks. Such a structure change affected molecular vibration and finally to photoemission spectrum in infrared region. The dication-$C_{23}H_{12}$ showed featured bands at 3.2, 6.3, 7.7, 8.6, 11.2, and 12.7 micrometer. It was surprising that those calculated bands coincided well with astronomically observed bands in many planetary nebulae. To confirm our study, large size molecule of $C_{53}H_{18}$ was studied. Calculation reproduced again similar astronomical bands. Also, small size molecule of $C_{12}H_8$ showed good coincidence with the spectrum observed for young stars. This paper would be the first report to indicate the specific PAH in space.

**Key words**: PAH, void, spin state, DFT, planetary nebula, infrared spectrum

## 1. Introduction

Graphene and graphite like carbon materials are candidates for showing carbon based ferromagnetism[1)-6)]. There are many capable explanations based on impurities[7)], edge irregularities[8)-10)] or defects[11)-16)]. However, origin of magnetic ordering could not be thoroughly understood. We previously reported in the same series paper[17)] that void defect in graphene nano ribbon (GNR) induces highly spin polarized magnetism investigated by density functional theory (DFT). Calculated result showed good coincidence with experiments[18)-20)]. Also, in our recent series paper[21)] we applied to graphene molecule. Void induced molecule was deformed to a featured structure having one carbon pentagon ring among hexagon networks. It was a surprise that most of graphene molecules show magnetism with the stable spin state of triplet, not singlet. Unfortunately, on laboratory experiment, such small pure carbon molecule did not show any magnetic feature. The reason may be molecule-to-molecule interaction at high molecular density conditions on earth of $10^{10} \sim 10^{23}$ molecules/cm³, which may bring paramagnetic canceling. We tried to look at astronomical carbon dust floating in interstellar and circumstellar space under ultra-low-density condition of 1~100 molecules/cm³. It is almost isolate molecule. While DFT calculation gives solution on isolate molecule. It was a surprise that calculated infrared spectrum showed good coincide with astronomically observed infrared spectrum, especially spectra of carbon rich planetary nebulae[22)23)].

In this paper, we like to try the case of hydrogenated graphene molecule, which is named polycyclic aromatic hydrocarbon (PAH) in chemistry. It is well known that PAH does not show any magnetic feature, which is typical diamagnetic material. Our question is that such non-magnetic property is common even for a case of void-defect induced PAH. Here again, we like to compare DFT calculated infrared spectrum with astronomically observed one. Molecular structure affects to molecular vibration, finally to infrared spectrum. The interstellar gas and dust show featured mid-infrared emission at 3.3, 6.2, 7.6, 7.8, 8.6, 11.2, and 12.7 μm, which are ubiquitous peaks observed at many astronomical objects[22)-29)]. Current common understanding is that these astronomical spectra come from the vibrational modes of PAH. There are many laboratory spectroscopy data[30)-33)] and DFT analysis[34)-39)]. However, despite long-term efforts, until now there is not any identified specific PAH. In this study, we will indicate unexpected coincidence of calculated emission spectrum of specific void induced PAH with above

---

Corresponding author: Norio Ota (n-otajitaku@nifty.com).

ubiquitously observed spectrum. In addition, to confirm our finding, larger size molecule and smaller one will be compared.

## 2. Model Molecules and Calculation Method

### 2.1 Model molecules

Model molecules are illustrated in Fig. 1. Starting mother molecule is typical PAH of coronene-$(C_{24}H_{12})$ having seven carbon hexagon rings. In this paper, we apply one assumption of single void-defect on initial molecule. On laboratory experiment, void will be created by an attack of high-speed particle as like proton or Argon. In interstellar and circumstellar space, high speed cosmic ray, mainly proton and electron, may attack on PAH. As illustrated on top of Fig. 1, high speed particle attacks mother molecule $(C_{24}H_{12})$ and kick out one carbon atom. Deformed molecular structure depends on void position as marked by red letters of c, d, and e. Void induced molecule is also named by suffixing as $(C_{23}H_{12}$-c), $(C_{23}H_{12}$-d) and $(C_{23}H_{12}$-e). DFT calculation resulted that there occurs serious structure change as shown in columns of Fig. 1. In case of $(C_{23}H_{12}$-c), two carbon pentagon rings are created among hexagon-ring network. Side view shows Y-shaped configuration. In case of $(C_{23}H_{12}$-d), one pentagon ring is created, where one extra hydrogen atom bonded with a carbon. Molecular structure is umbrella like curved one. Fragment of (C-H) will be kicked out in case of void-e to induce $(C_{23}H_{11}$-e) having one pentagon ring on a flat molecule. In this study, we suppose isolate molecule, which means no molecule-to-molecule interaction, and no energy competition between species. All species will be realized. To find size dependence, we will add larger molecule of $(C_{54}H_{18})$ in section 6, and smaller one of $(C_{13}H_{9})$ in section 7.

### 2.2 Calculation Methods

In calculation, we used DFT[40, 41] with the unrestricted B3LYP functional[42]. We utilized the Gaussian09 software package[43] employing an atomic orbital 6-31G basis set[44]. Unrestricted DFT calculation was done to have the spin dependent atomic structure. The required convergence of the root-mean-square density matrix was $10^{-8}$. Based on such optimized molecular configuration, fundamental vibrational modes were calculated, such as C-H and C-C stretching modes, C-H bending modes and so on, using the same Gaussian09 software package. This calculation also gives harmonic vibrational frequency and intensity in infrared region. The standard scaling is applied to the frequencies by employing a scale factor of 0.965 for PAH from the laboratory experimental value on coronene $(C_{24}H_{12})$[45]. Correction due to anharmonicity was not applied to avoid uncertain fitting parameters. To each spectral line, we assigned a Gaussian profile with a full width at half maximum (FWHM) of 4cm$^{-1}$.

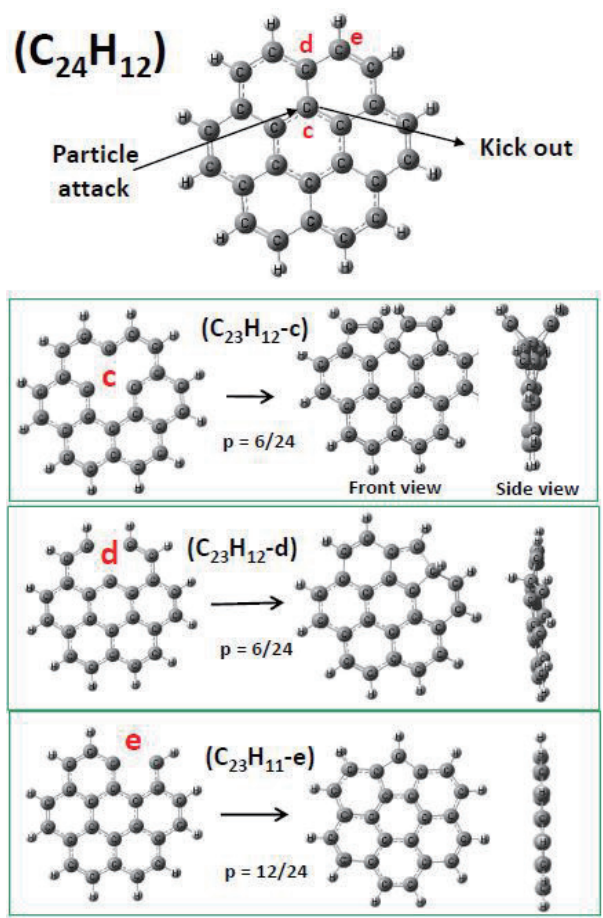

**Fig. 1** Void defect on mother molecule $(C_{24}H_{12})$. Deformed structure depends on void position suffixed by c, d, and e.

## 3. Spin State Analysis

Spin-multiplicity was studied for charge neutral molecules. In this study, we dealt total molecular spin $\mathbf{S}$ (vector). Molecule is rotatable material, easily follows to the external magnetic field of z-direction. Projected component to z-direction is maximum value of $S_z$, which becomes a good quantum parameter. In molecular magnetism, $S_z=2/2$ is named as triplet spin-state. Example is shown in Fig. 2 for void-c type molecule. As shown in (A), discussed in our previous paper[21], pure carbon molecule of $(C_{23}$-c) has 3 radical carbons holding 6 spins as initial void. One radical carbon holds two spins, which are forced to be parallel up-up spins (by red arrows in Fig. 2) or down-down spins (blue) for avoiding large coulomb repulsion due to Hund's rule[46]. Capable spin-states are $S_z=0/2, 2/2, 4/2$ and $6/2$. Three couples of spin-pair will be partially cancelled and remain one pair to be triplet. The most stable spin-state was $S_z=2/2$, which is 0.49 eV lower energy than that of $S_z=0/2$. Other spin states of $S_z=4/2$, and $6/2$ were rather unstable. DFT calculated spin cloud of $S_z=2/2$ is mapped on right showing up-spin by red and down-spin by blue.

Here, important question is that such magnetism is the same or not, even the case of void induced PAH as like ($C_{23}H_{12}$-c). It was interesting that calculated result was contrary to pure carbon case. Resulted spin configuration was illustrated in (B) of Fig. 2. The most stable spin state was $S_z=0/2$ to be nonmagnetic. Calculated energy of $S_z=2/2$ was 1.2eV higher (unstable) than that of $S_z=0/2$. Other void position cases also show that void-defect induced PAH have stable spin-state of $S_z=0/2$.

Hydrogen played an important role to diminish magnetism by bringing SP3-bond among SP2-network. Detailed molecular structure of ($C_{23}H_{12}$-c) is shown in Fig. 3. There is one SP3 bond marked by blue circle. Four carbons (1C, 2C, 3C and 4C) are bonded to a center carbon (0C). Six electrons in an initial void are used to make such SP3 bond and to diminish initial paired spins. There remains no spin-pair. Coulomb repulsive force between two hydrogen atoms (+0.14e for 1H and 2H) place them as far as possible, while carbon (0C) has negative charge of -0.26e, which attracts above hydrogen atoms. These forces bring SP3-bond.

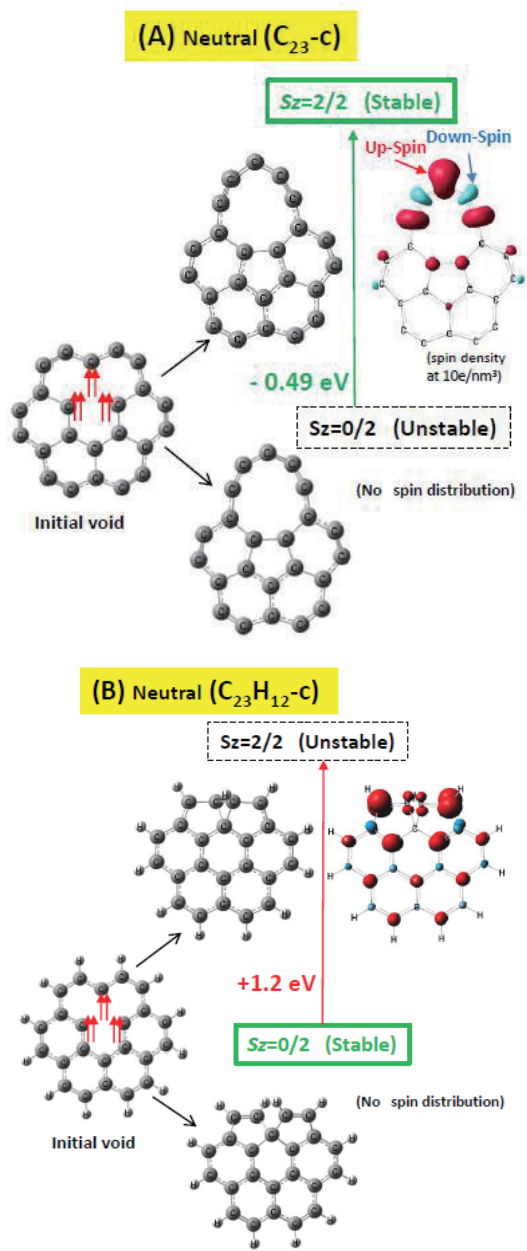

**Fig. 2** Stable spin state of void induced pure carbon ($C_{23}$-c) is triplet as shown in (A), while singlet for hydrocarbon ($C_{23}H_{12}$-c) in (B).

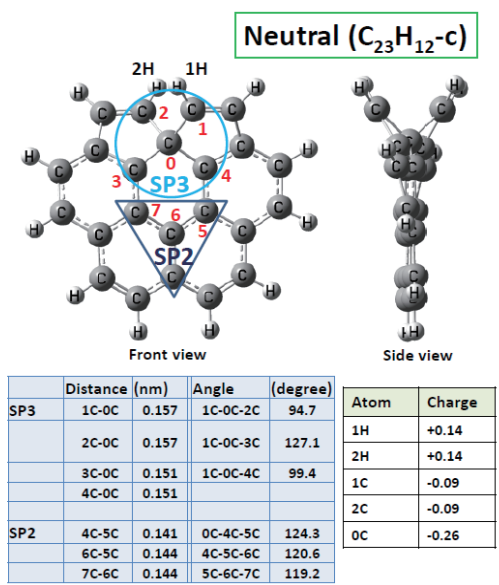

**Fig. 3** Molecular structure of neutral ($C_{23}H_{12}$-c).

### 4. Ionized Molecule and Infrared Spectrum

It is well known that PAH species having odd number molecular electrons show paramagnetic feature. Typical example is ionized PAH's. In case of ($C_{23}H_{12}$-c), energy diagram for ionization was illustrated on left of Fig. 4. It takes 6.50 eV to extract one electron from neutral molecule, which induce mono-cation ($C_{23}H_{12}$-c)[1+]. Similarly, di-cation needs additional 10.66 eV. On laboratory, high energy photo-illumination can realize such ionization. In interstellar space, central star can do such illumination on cosmic dust [29]. By photoionization, angle of 1C-0C-2C varies from 94.7 to 95.4 degree. Permanent dipole moment D increases from 0.65 to 1.20 Debye. Such structure change affects molecular vibration and infrared spectrum.

As illustrated on top of Fig. 5, high energy photon kicks out one electron of the highest occupied molecular orbit (HOMO). Remained one electron give rise to mono cation with spin-state of $S_z=1/2$. Spin distribution was calculated as shown on bottom left, which comes from the difference of total up-spins and total down-spins. HOMO is a major part of remained spin, of which orbital was illustrated on bottom right, where green cloud is positive sign wave-function and dark red negative one.

Calculated infrared spectra are shown in Fig. 6. Molecules are mother molecule ($C_{24}H_{12}$), void induced molecules of ($C_{23}H_{12}$-c), ($C_{23}H_{12}$-d) and ($C_{23}H_{11}$-e). It should be noted that infrared spectrum is sensitive to ionization. Neutral molecule showed single main band, whereas mono-cation three or more major bands, and di-cation more complex one.

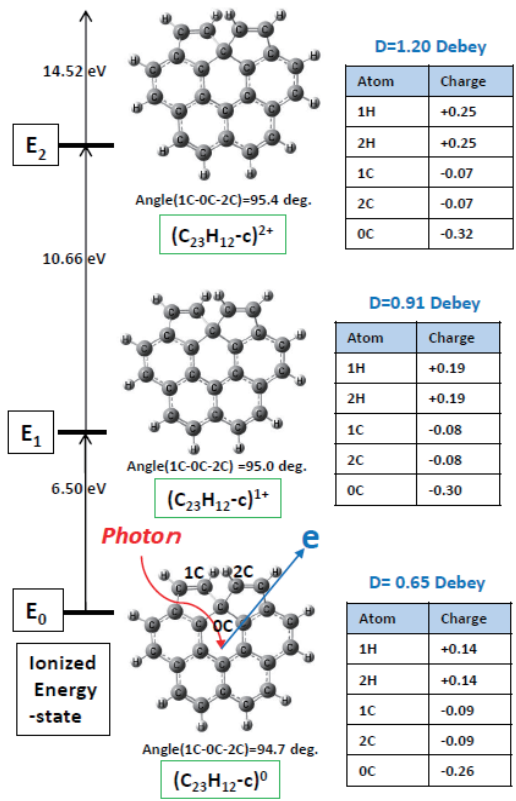

**Fig. 4** Energy diagram of ionized ($C_{23}H_{12}$-c).

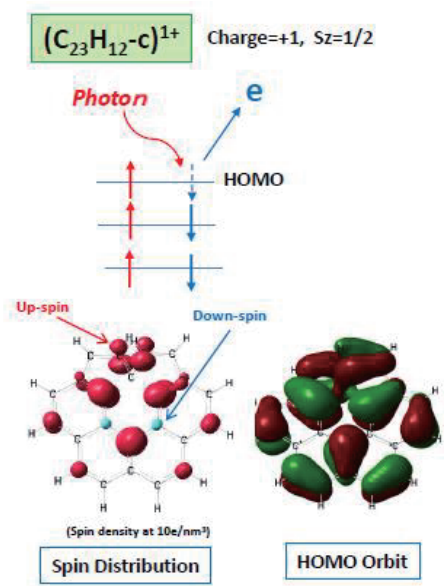

**Fig. 5** Spin distribution and HOMO orbit of ($C_{23}H_{12}$-c)$^{1+}$.

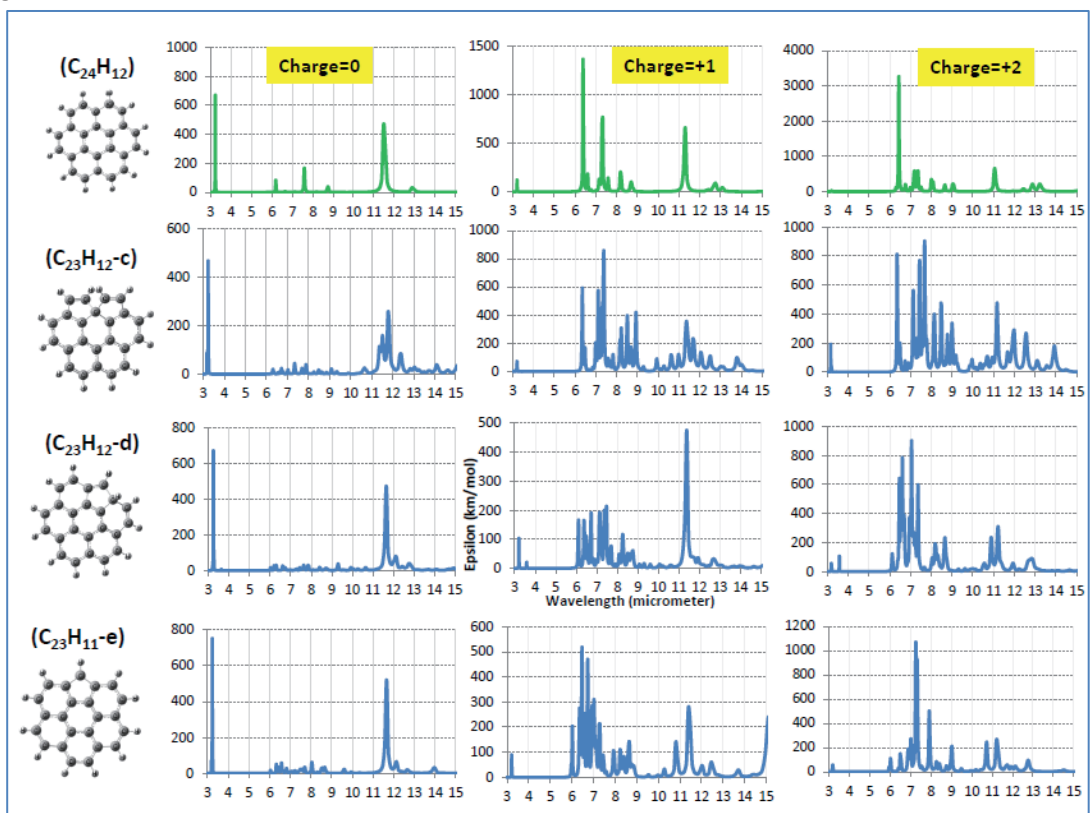

**Fig. 6** Calculated infrared spectra of ($C_{24}H_{12}$) family.

## 5. Comparison with Astronomical Observation

### 5.1 Infrared spectrum of planetary nebula

Vacuum level on earth is in a range of $10^9$-$10^{14}$ molecules/cm$^3$. While in interstellar and circumstellar space, molecular density will be 1-100 molecules/cm$^3$. Molecule has almost no interaction with other molecules. Additionally, cosmic dust will be kept under temperature less than 10K. Those conditions are favorable for comparing DFT calculation, because that DFT essentially gives isolate and temperature-zero solution.

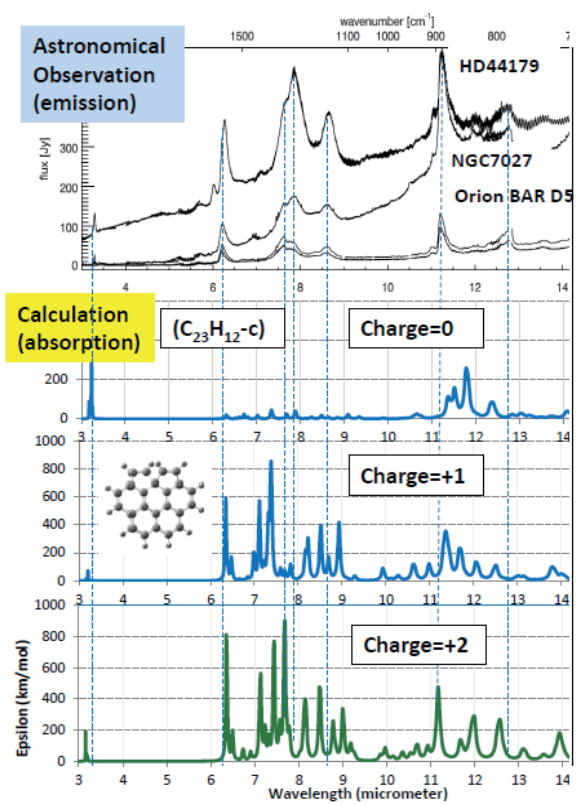

**Fig. 7** Calculated infrared spectra compared with astronomically well observed bands. We can see good coincidence with di-cation $(C_{23}H_{12}\text{-}c)^{2+}$.

On top of Fig. 7, astronomical spectra are illustrated for cases of planetary nebulae HD44179, NGC7027, and Orion Bar D5, reported by Boersma et al.[47]. They are few thousand light-years away, very far distance each other. It should be noted that those spectra show common infrared bands at 3.3, 6.2, 7.6, 7.8, 8.6, 11.2, and 12.7 μm. Such observation suggests the common essential mechanism on the creation and evolution of PAH in space. DFT calculated spectra listed in Fig. 6 were compared with such observed spectra. Among them, di-cation $(C_{23}H_{12}\text{-}c)^{2+}$ show major bands at 3.2, 6.3, 7.6, 7.8, 8.6, 11.2, and 12.6 μm. It was amazing that $(C_{23}H_{12}\text{-}c)^{2+}$ could well reproduce observed one as shown on bottom of Fig. 7. This may be first indication to suggest the existence of specific PAH in space. Despite over 30 years many efforts, until now there is not any identified specific PAH. In this study, we like to check whether our finding is an accidental coincidence or reasonable one by trying larger and smaller molecules in later sections.

### 5.2 Emission spectrum and fundamental mode

It should be noted that the astronomically observed spectra are seen in emission. A central star of nebula may illuminate cosmic molecules and excites them to give rise to infrared emission. Detailed discussion was reported by Li and Drain[48),49)]. Emission calculation on $(C_{23}H_{12}\text{-}c)^{2+}$ was done by Dr. Christiaan Boersma, NASA Ames laboratory, based on our fundamental vibrational mode analysis in private communication in 2014. He tried emission calculation supposing 6eV photoexcitation, scaling factor of 0.958, and FWHM of 15cm$^{-1}$. Result is shown in Fig. 8 by red on bottom compared with his observed data on NGC7023 nebula on top. Again, it was a surprise that emission calculation shows good coincidence with observation. Also, we regard that, in case of sufficient large photoexcitation, calculated absorption spectrum is a mirror image of emission one due to the theory of Einstein's emission coefficient[48),49)]. In this study, we like to compare astronomical spectrum simply with DFT calculated spectra.

Fundamental mode was analyzed on Table 1. There are 99 modes for 35 atoms. Zero-point vibrational energy is 7.44 eV above total electronic energy. The highest vibrational energy of mode-99 is 3284 cm$^{-1}$ (=0.407 eV) above zero-point vibrational energy. Mode-98 corresponds to observed 3.3 μm band, which comes from C-H stretching at carbon pentagon sites. Similarly, mode-87 by C-C stretching at hexagon sites corresponds to observed 6.2 μm band, also mode-71 by C-H in-plane bending and C-C stretching to 7.6 μm observed band, mode-69 to 7.8 μm observed band. Both mode61 and 62 may contribute to 8.6 μm observed band. Mode-43 was featured by C-H out-of-plane bending at all outer carbon sites, correspond to 11.2 μm observed band.

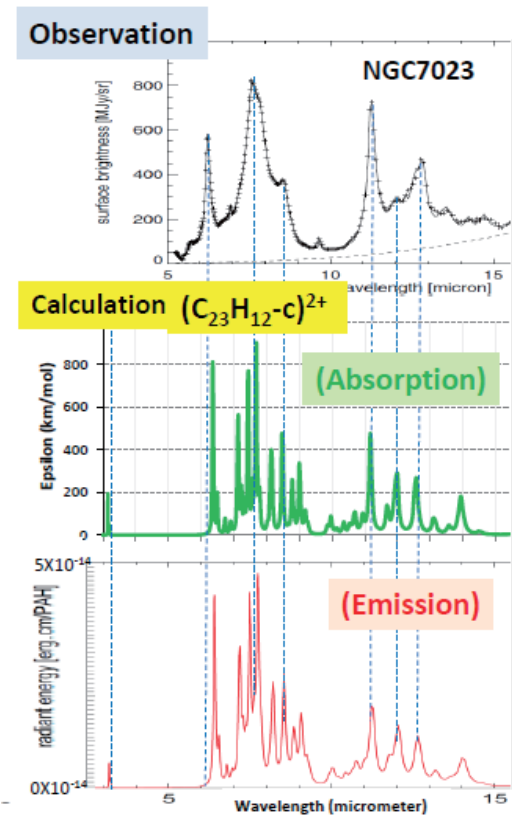

**Fig. 8** Calculated emission spectrum on bottom was compared with observed one of NGC7023 nebula on top. Absorbed bands in middle are mirror image of emission one.

**Table 1** Fundamental mode of $(C_{23}H_{12}\text{-}c)^{2+}$.

| Mode number | Energy (cm$^{-1}$) | Wavelength after scaling (μm) | Vibrational Intensity (km/mol) | Vibrational behavior |
|---|---|---|---|---|
| 1 | 97.0 | 106.9 | 4.6 | |
| 11 | 352.3 | 29.4 | 86.9 | molecular out-of-plane vibration |
| 17 | 487.0 | 21.3 | 46.5 | molecular out-of-plane vibration |
| 25 | 626.4 | 16.5 | 45.5 | molecular out-of-plane vibration |
| 36 | 825.6 | 12.6 | 81.0 | C-H out-of-plane bending at pentagon site |
| 43 | 927.8 | 11.2 | 136.1 | C-H out-of-plane bending at all sites |
| 56 | 1150.7 | 9.0 | 101.2 | C-H in-plane bending at all sites |
| 61 | 1221.9 | 8.5 | 165.8 | C-H in-plane bending, C-C stretching |
| 62 | 1270.2 | 8.2 | 103.4 | C-H in-plane bending, C-C stretching |
| 69 | 1349.8 | 7.7 | 290.1 | C-H in-plane bending, C-C stretching |
| 70 | 1369.2 | 7.6 | 61.9 | C-H in-plane bending, C-C stretching |
| 71 | 1393.7 | 7.4 | 258.5 | C-H in-plane bending, C-C stretching |
| 74 | 1433.1 | 7.2 | 57.4 | C-H in-plane bending, C-C stretching |
| 75 | 1451.7 | 7.1 | 102.5 | C-H in-plane bending, C-C stretching |
| 77 | 1457.2 | 7.1 | 84.6 | C-H in-plane bending, C-C stretching |
| 86 | 1625.3 | 6.4 | 60.7 | C-C stretching at hexagons |
| 87 | 1630.6 | 6.4 | 239.4 | C-C stretching at hexagons |
| 98 | 3283.5 | 3.2 | 42.6 | C-H stretching at pentagon sites |
| 99 | 3284.0 | 3.2 | 13.0 | |

## 6. Large Molecule ($C_{53}H_{18}$)

### 6.1 Model molecules

Possible origin of featured spectrum was suggested to be SP3 defect among SP2 network. Such mechanism should not depend on molecular size. Here, we tried large molecule starting from circum-coronene ($C_{54}H_{18}$). As shown on top of Fig. 9, there are 6 kind of void positions named by a, b, c, d, e, and f.

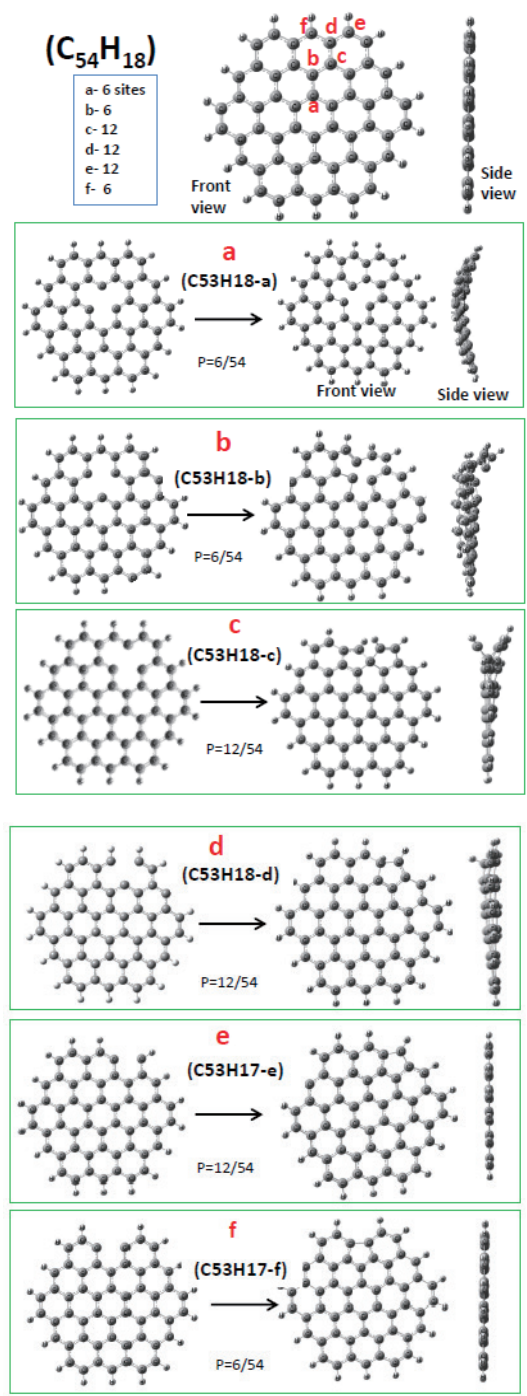

**Fig. 9** Void-defect induced species of ($C_{54}H_{18}$) family.

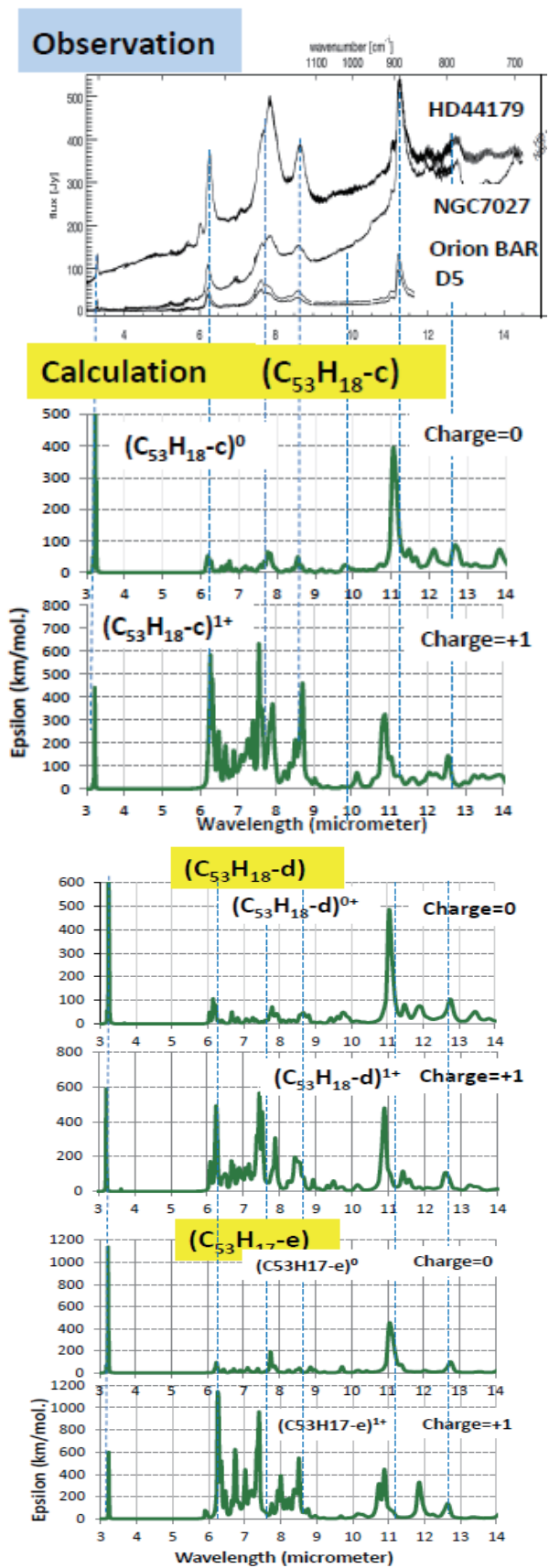

**Fig. 10** Calculated spectra of ($C_{53}H_{18}$) species compared with astronomical observation.

Void-a induces a species of ($C_{53}H_{18}$-a) showing an umbrella like structure having one carbon pentagon ring as illustrated in column-a of Fig. 9. Also, ($C_{53}H_{18}$-b) show a similar structure. It is interesting that ($C_{53}H_{18}$-c) has two pentagon rings and show similar structure to previously calculated case of ($C_{23}H_{12}$-c). In ($C_{53}H_{18}$-d), one hydrogen brings an umbrella like structure introduced by SP3-H bond. In case of ($C_{53}H_{17}$-e), C-H fragment was removed, which show flat structure including one pentagon ring. Also, void-f brings a flat molecule of ($C_{53}H_{18}$-f).

### 6.2 Infrared spectrum

Infrared spectra for charge 0 and +1 of void-c, -d and –e induced species are calculated as shown in Fig. 10. Di-cation spectra were listed on Appendix. Among them, it was amazing that mono-cation molecule of ($C_{53}H_{18}$-c)$^{1+}$ shows good coincidence with astronomically observed bands at 3.3, 6.2, 7.6, 7.8, and 8.6 μm. Whereas, neutral molecule shows bands at 11.1 and 12.7μm close to observation. Observed spectrum will be a sum of those neutral and ionized molecule's spectrum. Similarly, ($C_{53}H_{18}$-d) also shows good coincidence with observation. In addition, ($C_{53}H_{17}$-e) shows coincidence at 6.2, 7.6, and 8.6 μm with mono-cation, and at 11.1 μm with neutral one. It was concluded that many species could reproduce ubiquitously observed astronomical spectrum. It was revealed that our finding is not an accidental one.

## 7. Small Molecule ($C_{12}H_8$)

### 7.1 Model Molecules

We tried small size molecules of ($C_{12}H_9$) and ($C_{12}H_8$) created from mother molecule of ($C_{13}H_9$) as shown in Fig. 11. Void-c induced molecule of ($C_{12}H_9$-c) has a complex structure with void creation capability of p=1/13 (1 void among 13 carbons). Void-d induces umbrella like configuration of ($C_{12}H_9$-d). Void-e and void-f both lose one (C-H) fragment to show the same flat configuration as ($C_{12}H_8$-e), and ($C_{12}H_8$-f). This is a major species having void capability of p=9/13.

### 7.2 Infrared Spectrum

Calculated spectra are listed in Fig. 12, which show different spectra with ubiquitously observed one as illustrated in Fig. 7. It was amazing that some coincidence of calculated spectra was found with young star's observed spectrum, especially observed at protoplanetary disks around the Herbig Ae/Be and T Tauri stars, which were reported by Acke et al. in 2010[50] and by Seok and Li in 2017[51].

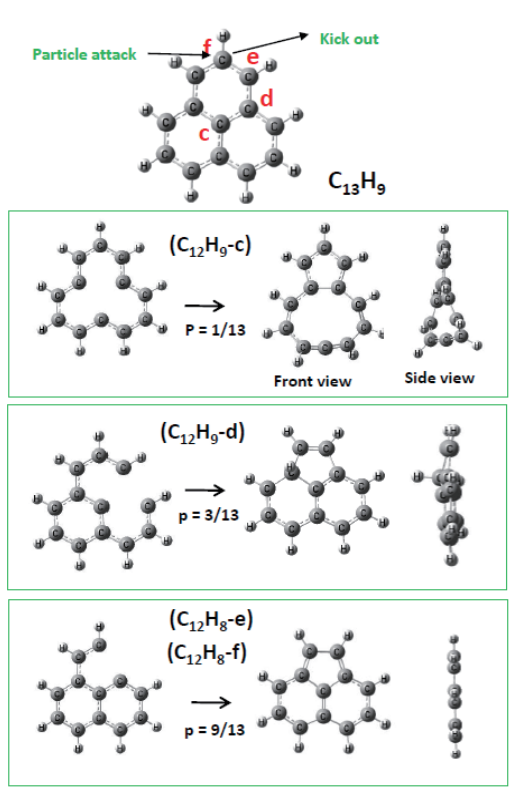

Fig 11 Void induced small size molecules.

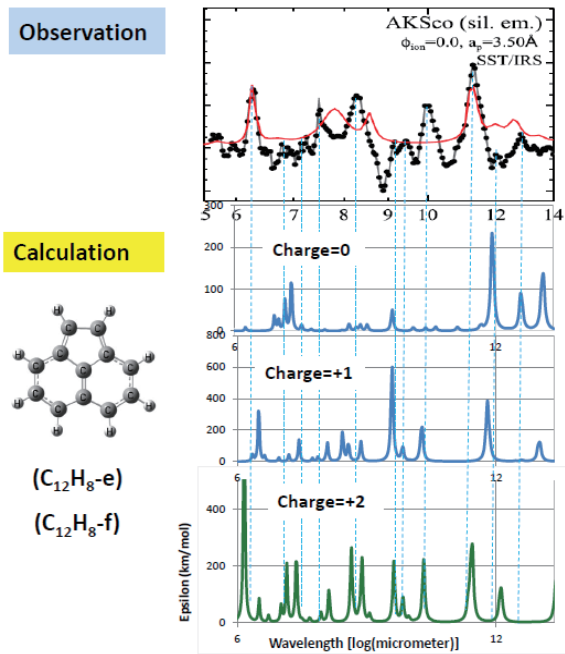

Fig. 13 Observed spectrum of young star AKSco was well reproduced by $(C_{12}H_8\text{-e})^{2+}$.

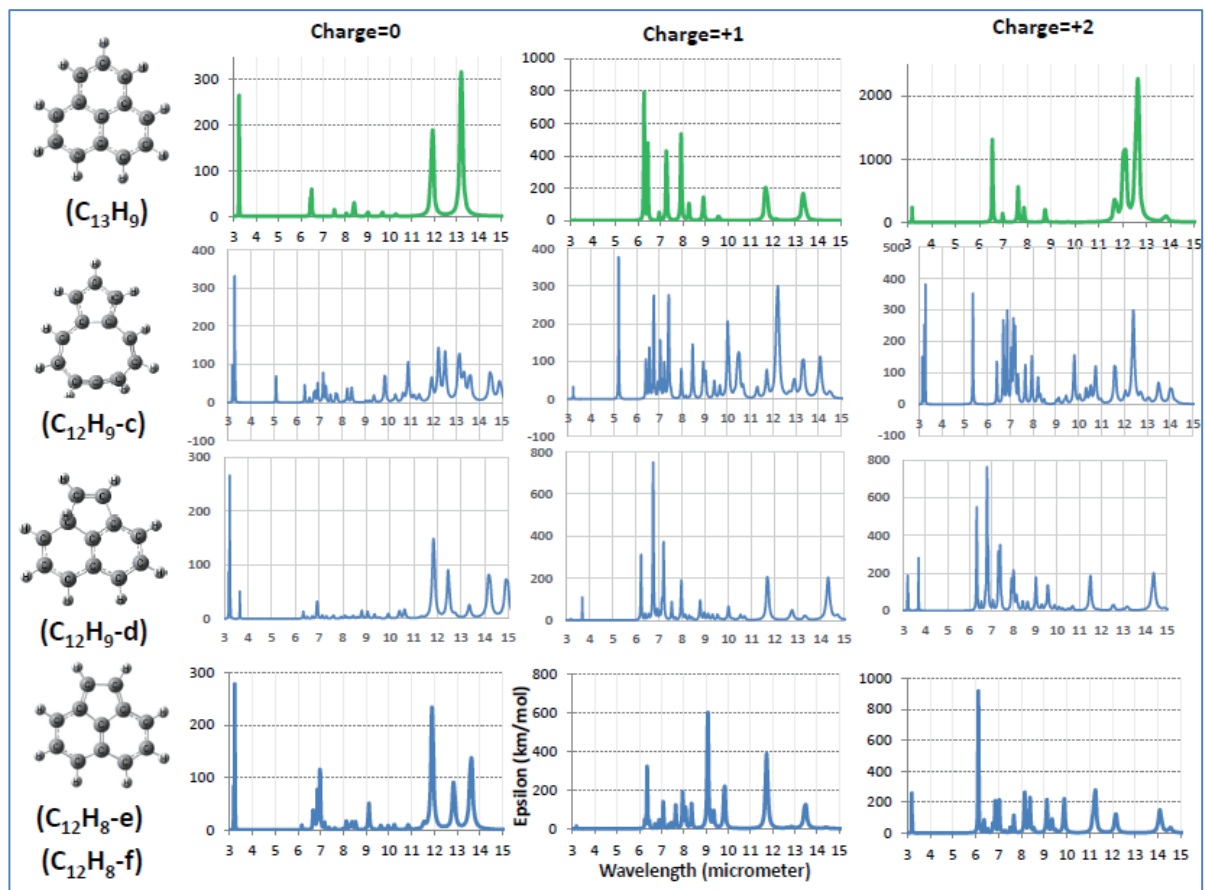

Fig. 12 Small molecule's calculated infrared spectra.

Recently, young star interested many scientists, because it is an analogy of baby age of our solar system and planets. We can understand how planet system will be created in the Universe. Among more than 60 observed spectra by Seok and Li[51], we found 9 samples coincident with spectrum of di-cation $(C_{12}H_8\text{-e})^{2+}$. Typical example is AKSco as shown in Fig. 13. It was a surprise that such complex spectrum can be reproduced well with calculated spectrum of di-cation $(C_{12}H_8\text{-e})^{2+}$. We can see good coincidence at many bands of 6.2, 6.8, 7.2, 7.5, 8.2, 9.1, 10.0, 11.3, 12.0, and 12.7 μm. Another example is shown in Fig. 14 for observed spectrum of HD31648. Again, there are many complex bands. We analyzed that most bands coincide well again with calculated one of $(C_{12}H_8)^{2+}$, which are marked by blue dotted lines. Also, we found that several bands will be identified partly by $(C_{23}H_{12}\text{-c})^{2+}$ marked by red dotted lines. We could reproduce observed spectrum by a sum of those molecules.

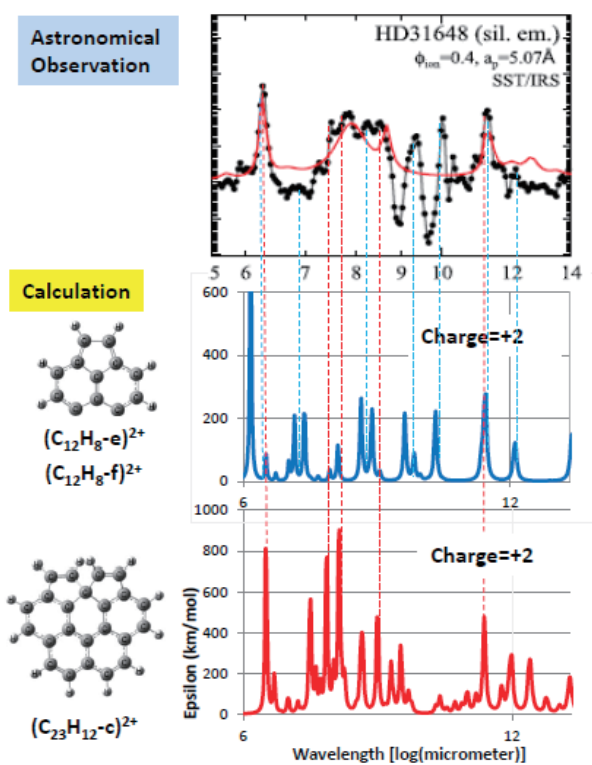

**Fig. 14** Complex spectrum of young star HD31648 was reproduced well by a sum of $(C_{12}H_8\text{-e})^{2+}$ and $(C_{23}H_{12}\text{-c})^{2+}$.

## 8. Conclusion

Void-defect induced magnetism and structure change of polycyclic aromatic hydrocarbon molecules (PAH) were studied by density functional theory (DFT) and by astronomical observation.
(1) Model molecule of $(C_{23}H_{12})$ was introduced by making a void-defect on $(C_{24}H_{12})$. Induced species have one or two carbon pentagon rings among hexagon ring networks.
(2) Single void holds six spins and cause spin multiplicity. Stable spin-state of $(C_{23}H_{12})$ was singlet, which is contrary to pure carbon case of $(C_{23})$ with triplet one. Hydrogen plays an important role to diminish magnetism by bringing SP3-bond among SP2-networks.
(3) Molecular charge brings serious change on magnetism and structure, which finally affect molecular vibrational spectrum in infrared region. Di-cation molecule $(C_{23}H_{12})^{2+}$ shows featured bands at 3.3, 6.2, 7.6, 7.8, 8.6, 11.2, and 12.7 μm. It was amazing that those calculated bands coincident well with astronomically observed bands. This study will be the first indication to suggest specific PAH in space.
(4) To confirm our finding, large model molecule of $(C_{53}H_{18})$ was tested. Most of mono-cation $(C_{53}H_{18})^{1+}$ showed again featured bands reproducing astronomically observed one.
(5) In addition, small size molecules of $(C_{12}H_9)$ are tested. It was a surprise that calculated spectrum shows good coincidence with young star's unusual complex spectrum.

It was revealed that void induced PAH is a promising candidate of cosmic hydrocarbon molecule.

## Acknowledgement

Aigen Li is supported in part by NSF AST-1311804 and NASA NNX14AF68G.

Norio Ota would like to say great thanks to Dr. Christiaan Boersma, NASA Ames Research Center, to provide the calculated emission infrared spectrum based on our DFT analysis, and to apply it to his observation.## References

**Appendix** Calculated spectrum of (C$_{53}$H$_{18}$) species.

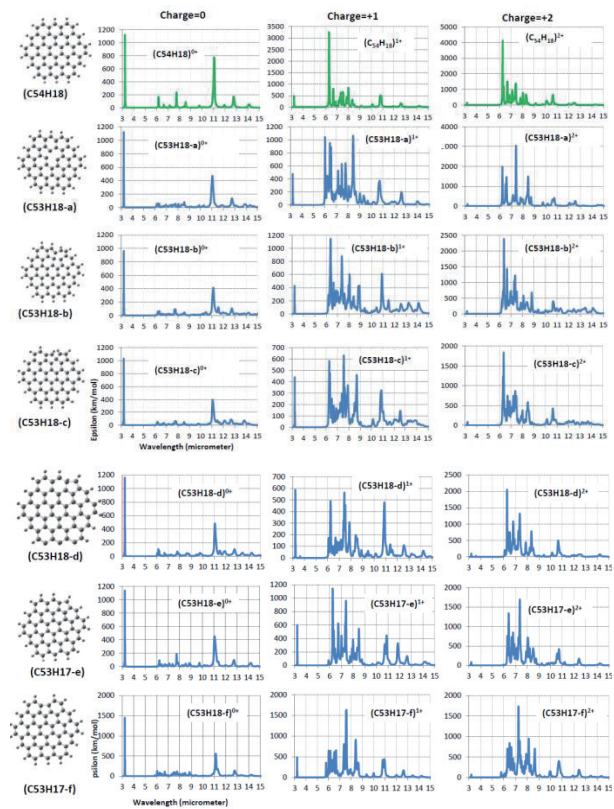